\documentclass[twocolumn]{article}
\usepackage{microtype}
\usepackage{float}
\usepackage{authblk}
\usepackage{graphicx}
\usepackage{wrapfig}
\usepackage{braket}
\usepackage{amsmath}
\usepackage{amssymb}
\usepackage{amsfonts,amsxtra}
\usepackage{algorithm}
\usepackage[noend]{algpseudocode}
\algrenewcommand{\algorithmiccomment}[1]{\hfill\parbox[t]{0.4\linewidth}{\raggedright$\triangleright$ #1}}

\usepackage{natbib}
 \bibpunct[, ]{(}{)}{,}{a}{}{,}%
\usepackage[
  top=2.5cm,
  bottom=2.5cm,
  left=2.5cm,
  right=2.5cm
]{geometry}
\usepackage{xcolor}
\usepackage[most]{tcolorbox}
\usepackage{hyperref}
\hypersetup{
    colorlinks=true,
    linkcolor=blue,
    citecolor=blue,
    urlcolor=blue
    }
\usepackage{subcaption}
\usepackage{parskip}
\usepackage{indentfirst}
\title{Quantum-enhanced Monte Carlo Tree Search framework for combinatorial optimization problems}
\author[1]{Yohan Finet\thanks{yohan.finet@usherbrooke.ca}}
\author[1]{Yves Bérubé-Lauzière\thanks{yves.berube-lauziere@usherbrooke.ca}}
\author[1]{Victor Drouin-Touchette\thanks{victor.drouin-touchette@usherbrooke.ca}}
\affil[1]{Institut Quantique and Département de Génie Électrique et de Génie Informatique, Faculté de Génie, Université de Sherbrooke, Sherbrooke, Québec, J1K 2R1, Canada}
\date{(Dated: \today)}

\begin{document}

\twocolumn[
\maketitle

\begin{center}
\begin{minipage}{0.85\textwidth}
\setlength{\parindent}{1.25em}

Over the past decades, the operations research community has developed numerous effective optimization algorithms, yet quantum computing is emerging as a new computational paradigm with the potential to approach optimization problems more efficiently. Grover's algorithm offers a provable speedup for combinatorial optimization, but its circuit depth places it beyond current noisy intermediate-scale quantum (NISQ) devices. A more accessible alternative is to reformulate the optimization problem as a quadratic unconstrained binary optimization (QUBO) problem and apply quantum annealing; however, practical problem instances remain out of reach for existing hardware. We introduce \textsc{AtomTreeSearch}, a hybrid classical-quantum algorithm that integrates a quantum subroutine natively implementable on neutral-atom quantum computers within a Monte Carlo Tree Search framework. At each expansion step, a maximal weighted independent set of candidate actions provided by the quantum processor is selected, and these collective actions are performed to obtain a child node. We benchmark our method on the Traveling Salesman Problem, with instances of up to 60 cities on random Euclidean instances and up to 100 cities on TSPLIB instances. Our hybrid algorithm generally matches or outperforms both \texttt{OR-Tools} and simulated annealing on these instances, and we find that the quantum subroutine produces more diverse and higher-quality branches compared to classical alternate subroutines. These results suggest that carefully scoped quantum subroutines embedded in classical search frameworks represent a promising path toward near-term quantum utility in combinatorial optimization.

\end{minipage}
\end{center}

\vspace{2em}
]

{\renewcommand{\thefootnote}{\fnsymbol{footnote}}
\footnotetext[1]{yohan.finet@usherbrooke.ca}
\footnotetext[2]{yves.berube-lauziere@usherbrooke.ca}
\footnotetext[3]{victor.drouin-touchette@usherbrooke.ca}}

\section{Introduction} \label{sec:intro}
Combinatorial optimization (CO) problems are ubiquitous in industry, with applications in transportation, logistics, energy, communications, and healthcare for example \citep{crainic2024}. Their prevalence has motivated decades of research across multiple disciplines, most notably operations research (OR). While some CO problems can be solved efficiently (e.g., shortest path problems), many others are NP-hard \citep{Papadimitriou1982}. A broad class of such intractable problems admits natural formulations as integer linear programs (ILPs), which can be solved exactly using methods such as cutting planes, branch-and-bound, branch-and-cut, and column generation \citep{Desrosiers2024}. Although these approaches provide optimality guarantees, their worst-case exponential time complexity limits their practical applicability to large-scale instances. Metaheuristics offer a scalable alternative, but their effectiveness relies on problem-specific structural insights and careful empirical tuning, thereby requiring substantial domain expertise \citep{bengio2020}.

With the aim of reducing this need for expert knowledge, the machine learning (ML) community has brought forth a diverse set of approaches to address CO problems. Recent advances achieved with deep learning, a sub-field of ML focused on large parameterized function approximators, have demonstrated remarkable effectiveness in high-dimensional settings with abundant data. These developments suggest the potential to replace computationally intensive components of traditional OR algorithms with efficient approximators, learned in a generic way \citep{bengio2020}. For instance, \cite{Lodi2017} provide a review of learning techniques applied to variable and node selection in branch-and-bound algorithms.

ML-based methods for CO can be categorized into three families: (i) end-to-end learning, where a neural network is trained to directly map problem instances to solutions, was pioneered in a supervised setting by \cite{Vinyals2015} with pointer networks which were later trained with reinforcement learning by \cite{bello2017}; (ii) algorithm configuration, where ML is used to parameterize an OR algorithm, as in the work of \cite{kruber2017}, which focused on predicting whether a Dantzig--Wolfe decomposition is likely to be beneficial for a given MILP instance; and (iii) decision assistance, where an OR algorithm repeatedly queries an ML model for lower-level decisions and can be simply illustrated by the work of \cite{dai2017}. ML-based methods can also be classified as construction methods, which iteratively build a solution (e.g., \cite{Vinyals2015}), or improvement methods, which iteratively refine an initial solution (e.g., \cite{Costa2020}). Of particular relevance to this work, Monte Carlo Tree Search (MCTS), famously employed in the AlphaGo system that defeated a world champion in the game of Go \citep{AlphaGo2016}, has been adapted to combinatorial optimization (CO) problems, both as a construction method \citep{xing2020} and as an improvement method \citep{Fu2021}, and demonstrated promising performance. Section \ref{sec:mcts} provides a detailed overview of MCTS and its ability to balance exploration of the solution space and exploitation of information gathered from previous exploration to guide future search. For broader surveys, see \cite{mazyavkina2020} for reinforcement learning approaches to CO problems and \cite{bengio2020} for a more general review of ML approaches to CO problems. Despite their promise, these methods share notable limitations: training data generation is a subtle art and strongly influences performance, generalization to instances differing in size or structure from the training set is typically poor, optimality guarantees are generally absent, and interpretability remains challenging.

In parallel, quantum computing is emerging as a novel computational paradigm in which information is encoded in quantum bits (qubits), capable of existing in superpositions of basis states, $\ket{\psi} = \alpha \ket{0} + \beta \ket{1}$ with $|\alpha|^2 + |\beta|^2 = 1$. This property gives rise to quantum superposition, entanglement and interference which can be harnessed to perform computations beyond classical capabilities. The evolution of a quantum state is governed by the Schrödinger equation, $\frac{d}{dt}\ket{\psi(t)} = -\frac{i}{\hbar} H \ket{\psi(t)}$, where the Hamiltonian $H$ encodes the total energy of the system and may be time-dependent. A quantum computation can be understood as the task of preparing an initial state, evolving it under a prescribed Hamiltonian, and performing a measurement that probabilistically collapses the quantum state to a classical outcome \citep{Nielsen2010}. Within this framework, quantum computing offers alternative strategies for CO problems. Notably, Grover's algorithm \citep{Grover1996, roland2002quantum} achieves a quadratic speedup over brute-force search for unstructured problems, and this speedup extends, via quantum amplitude amplification, to a large family of classical heuristics that exploit problem structure \citep{Brassard2002}. More recently, \cite{jordan2025} proposed the decoded quantum interferometry (DQI) algorithm as a candidate for exponential speedup on certain narrowly defined optimization problems such as max-LINSAT and the optimal polynomial intersection problem. However, realizing these approaches in practice demands a number of coherent operations well beyond the reach of current noisy intermediate-scale quantum (NISQ) devices \citep{Preskill2018}, which suffer from short coherence times, high gate error rates, limited qubit connectivity and no fault-tolerant error correction.

Many NP-complete and NP-hard problems admit reformulations as quadratic unconstrained binary optimization (QUBO) problems of the form $\min_{\mathbf{x}} \mathbf{x}^{\intercal}\mathbf{Q}\mathbf{x}$, wherein constraints are incorporated into the objective function through the introduction of slack variables and penalty terms weighted by Lagrange multipliers. Once cast in QUBO form, CO problems may be further reformulated as Ising models \citep{Lucas2014}, in which a spin $s_i \in \{-1, 1\}$ is associated with each binary decision variable $x_i \in \{0, 1\}$ via the mapping $x_i = (s_i + 1)/2$. Promoting the classical spin variables to Pauli $\hat{\sigma}_i^z$ operators acting on qubit $i$, the energy of each spin configuration (i.e. binary solution) is then encoded by the Hamiltonian
\begin{equation}
    H = -\sum_{i<j} J_{ij}\, \hat{\sigma}_i^z \hat{\sigma}_j^z - \sum_{i=1}^{N} h_i\, \hat{\sigma}_i^z.
\end{equation}
This structure is amenable to two prominent quantum heuristic approaches. Quantum annealing \citep{Kadowaki1998} takes as input a problem Hamiltonian $H_P$, whose ground state encodes the solution to the QUBO problem, and an initial Hamiltonian $H_0$ with an easy-to-prepare ground state. A common choice is $H_0 = \sum_{i=1}^{N} \hat{\sigma}_i^x$, where $\hat{\sigma}_i^x$ denotes the Pauli $\hat{\sigma}_x$ operator on qubit $i$, whose ground state is the $N$-qubit uniform superposition $\ket{+}^{\otimes N} = \frac{1}{2^{N/2}}(|0\rangle + |1\rangle)^{\otimes N}$. The system is initialized in the ground state of $H_0$ and slowly evolved toward $H_P$. By the adiabatic theorem \citep{Born1928}, a sufficiently slow evolution guarantees that the system remains in the instantaneous ground state, thereby yielding the solution at the end of the schedule. However, for hard Ising problems the time required to satisfy the adiabatic condition can scale exponentially with system size, which is then the worst case scenario. In practice, the evolution is carried out in finite time, and quantum annealing operates as a quantum heuristic \citep{abbas24} to obtain low-energy, near-optimal solutions, whose quality degrade as evolution becomes shorter. The quantum approximate optimization algorithm (QAOA) \citep{farhi2014} is a hybrid classical-quantum method that prepares a quantum state via alternating problem and mixer parameterized unitaries, with parameters tuned by a classical outer loop to minimize the expected energy of the prepared state. Both approaches face significant practical limitations: quantum annealing is hindered by hardware connectivity constraints that are often incompatible with the graph structure of the target QUBO, while QAOA suffers from degrading performance with circuit depth on NISQ devices and the inclusion of another hard parameter-tuning optimization problem. In view of this, it is likely that any near-term quantum utility in practical applications will require hybrid algorithms that seamlessly integrate quantum subroutines into efficient classical workflows, carefully leveraging the capabilities of current quantum hardware where they are most effective \citep{abbas24}.

\begin{figure*}
  \centering
  \includegraphics[width=\textwidth]{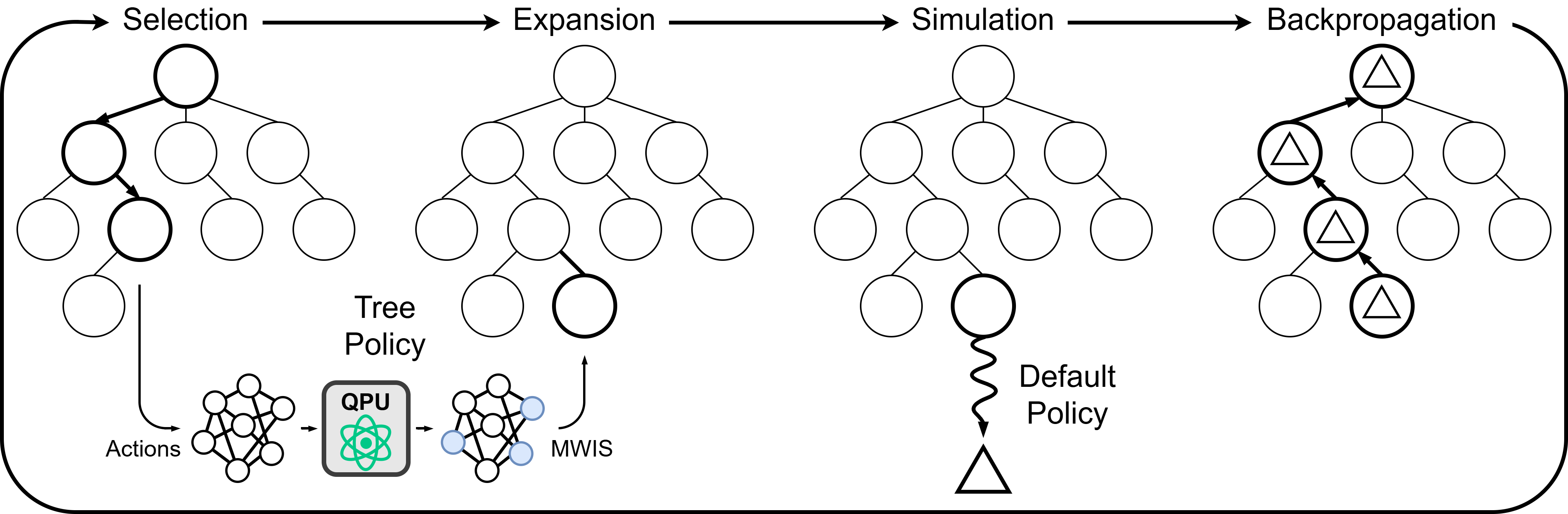}
  \caption{Illustration of \textsc{AtomTreeSearch} which consists of a MCTS algorithm where the expansion step selects a maximal weighted independent set of actions provided by a quantum processing unit (QPU) and performs these collective actions to obtain a child node.}
  \label{fig:mcts}
\end{figure*}

In this paper, we introduce \textsc{AtomTreeSearch}, a hybrid quantum-classical method illustrated in Fig.~\ref{fig:mcts}. The capability of neutral-atom quantum computers (the specificity of this architecture is discussed in detail in Section~\ref{sec:naqc}) to efficiently sample graph independent sets \citep{cazals2025_1, gibson2025quantum, perron2026iterative} is leveraged in a quantum subroutine within a classical MCTS framework, chosen for its inherent balance between exploration and exploitation. We formulate our algorithm for general combinatorial optimization problems and benchmark it on the paradigmatic traveling salesman problem (TSP). We further assess the impact of substituting the quantum subroutine with classical alternatives. We observe that using the quantum subroutine generally yields solutions on par with or superior to those obtained using classical alternatives. On random Euclidean TSP instances with up to 60 cities and TSPLIB \citep{reinelt1991} instances with up to 100 cities, our hybrid algorithm generally matches or outperforms both \texttt{OR-Tools} \citep{ortools} and a simulated annealing approach \citep{Kirkpatrick1983}. Our analysis suggests that the quantum subroutine facilitates the exploration of high-quality and diverse branches within the search tree, ultimately leading to improved solution quality. Our contribution shares similarity with other hybrid methods \citep{Montanaro_2020, Tran2021, Chao2023} and related works \citep{farhi1998, martonak2004, Sequeira2021, meyer2024}, though none combine these elements in the same way as \textsc{AtomTreeSearch}.

The remainder of this paper is organized as follows. In Section~\ref{sec:atomic_tree_search}, we introduce our quantum-enhanced MCTS framework for combinatorial optimization. Section~\ref{sec:TSP_app} provides an overview of the Traveling Salesman Problem, reviews the principal classical solution methods, and details how our framework can be adapted to address this problem. In Section~\ref{sec:results}, we present a comparative analysis of the results obtained with our algorithm and those achieved using classical approaches. Finally, we conclude in Section \ref{sec:conclusion} with a discussion of the broader applicability of our method to other combinatorial optimization problems, such as other routing problems, as well as potential improvements to both the framework and its implementation for the Traveling Salesman Problem.

\section{AtomTreeSearch}\label{sec:atomic_tree_search}

Before describing \textsc{AtomTreeSearch}, we first review the basics of Markov Decision Processes (MDPs) in Section~\ref{sec:mdp} and of MCTS algorithms in Section~\ref{sec:mcts}. We then present how we modify the usual Upper Confidence Bounds for Trees (UCT) algorithm by introducing a Maximum Weighted Independent Set (MWIS) problem in Section~\ref{sec:mwis_of_actions}. Finally, in Section~\ref{sec:naqc} we explain how a quantum subroutine running on a neutral-atom quantum computer can be used to sample solutions to this MWIS problem.

\subsection{Markov Decision Processes} \label{sec:mdp}

MDPs \citep{szepesvari2010, sutton2014} provide a formal framework for sequential decision-making. In this setting, actions are taken to interact with an environment and collect rewards. A state provides a complete description of the world. The set of all possible states is called the state space $S$, and the set of valid actions (or decisions) in state $s_t$ is called the action space $A(s_t)$. When an action $a_t$ is taken at time step $t$, the state transitions from $s_t$ to $s_{t+1}$ and a reward $r_{t+1}$ is received at the following time step. We say that an environment has the Markov property if its dynamics is defined by the probability distribution ${p(s' \mid s, a) = \Pr\{s_{t+1} = s' \mid s_t = s,\ a_t = a\}}$. In other words, the environment's response at time $t+1$ depends only on the state and action taken at time $t$. To formulate a CO problem as a MDP, states are defined to represent candidate solutions, while actions correspond to operations that map one solution to another.

The objective in the context of a MDP is to maximize the cumulative reward, or return, along a trajectory $\tau = (s_0, a_0, s_1, a_1, \dots)$ which is a sequence of states and actions. For a finite, undiscounted horizon of $T$ time steps, the return at time $t$ is given by ${R_t = \sum_{k=0}^{T-t-1} r_{t+k+1}}$, whereas for an infinite, discounted horizon it is defined as ${R_t = \sum_{k=0}^{\infty} \gamma^k r_{t+k+1}}$, where ${\gamma \in [0,1]}$ denotes the discount rate. This optimization is carried out through a policy $\pi$, i.e., a rule mapping states to actions. Policies may be deterministic, $a_t = \pi_\theta(s_t)$, or stochastic, $a_t \sim \pi_\theta(\cdot \mid s_t)$, and may be parameterized by $\theta$. The action-value function ${Q^\pi(s,a) = \mathbb{E}_\pi[R_t \mid s_t=s,a_t=a]}$ gives the expected return when starting in state $s$, taking action $a$, and then following policy $\pi$ indefinitely. MCTS algorithms, discussed in the next section, offer an approach to approximate the value of the actions that can be taken from a state.

\subsection{Upper Confidence Bounds for Trees} \label{sec:mcts}

Our method builds on the Upper Confidence Bounds for Trees (UCT) algorithm, a popular method within the MCTS family (for a review, see \cite{browne2012}). MCTS algorithms approximate the optimal policy for an MDP by iteratively constructing a partial search tree until some computational budget (e.g., time, memory, number of iterations) is reached. Each node in the tree corresponds to a state and edges linked to child nodes correspond to actions leading to subsequent states. One iteration proceeds in four phases, as illustrated in Fig. \ref{fig:mcts}:
\begin{enumerate}
    \item \textbf{Selection}: Starting from the root node, recursively select a child node to visit using a tree policy until reaching a node that is not fully expanded.
    \item \textbf{Expansion}: Add a new child node by executing an action not yet explored from the state of the selected node.
    \item \textbf{Simulation} (or rollout): From the new node, simulate a trajectory to a terminal state using a default policy (random or heuristic).
    \item \textbf{Backpropagation}: Propagate the obtained return back up the tree, updating the statistics (such as visit counts and value estimates) of each ancestor node along the path.
\end{enumerate}

A key characteristic of MCTS is its anytime property: the algorithm can be halted at any moment, yielding the best action to perform from the root node. In Alg. \ref{alg:mcts_search} we present the alternative formulation we use where the algorithm returns the best state encountered during the search, a notion that will be formally defined soon. Here, $s(v)$ denotes the state associated to node $v$. The $\textnormal{\textsc{Select}}$ function returns the node to be expanded, $\textnormal{\textsc{Expand}}$ creates and returns a new child node, and $\textnormal{\textsc{Terminal}}(s)$ indicates whether $s$ is terminal. A state is terminal when no actions can be performed from this state. Given the simplicity of the function $\textnormal{\textsc{Terminal}}$, we omit its pseudocode. Finally, $R$ is the return obtained upon reaching the terminal state after following the default policy and $a(\textnormal{\textsc{BestChild}}(v_0))$ is the action leading to the best child of the root node, where the meaning of ``best" depends on the implementation. $\textnormal{\textsc{Backpropagate}}$ simply implements the backpropagation phase (cf. Alg.~\ref{alg:mcts_backprop}).

As the notion of ``best state'' may be ambiguous in the general MDP setting, we formalize what we mean by this here. Along a given trajectory, the best state is defined as the one whose associated candidate solution attains the greatest (smallest) value of the objective function for a maximization (minimization) problem. Furthermore, if the candidate solution corresponding to the best state identified during the current MCTS iteration improves upon the best solution found in previous iterations, it replaces the latter as the incumbent best state.

\begin{algorithm}[t]
\caption{\textnormal{\textsc{MctsSearch}}($s_0, b$)}\label{alg:mcts_search}
\begin{tabular}[t]{@{}l@{\ }l@{}}
\textbf{Input:}  & The initial state $s_0$ and the computational \\
                 & budget $b$
\end{tabular}\\
\textbf{Output:} The best state found
\begin{algorithmic}[1]
\State create root node $v_0$ from $s_0$
\While{within $b$}
    \State $v \gets$ \Call{Select}{$v_0$}
    \If{$\neg$\Call{Terminal}{$s(v)$}}
        \State $v \gets$ \Call{Expand}{$v$}
    \EndIf
    \State $R \gets$ \Call{Simulate}{$v$}
    \State update best state if better state found
    \State \Call{Backpropagate}{$v,R$}
\EndWhile
\State \textbf{return} best state found
\Comment{Usually, return $a$(\Call{BestChild}{$v_0$}).}
\end{algorithmic}
\end{algorithm}

The selection phase plays a critical role in determining the structure of the tree generated by a given MCTS algorithm. To balance exploration (favoring infrequently visited nodes) and exploitation (favoring high-value nodes), UCT formulates the choice of child nodes as a multi-armed bandit problem \citep{Kocsis2006_1, Kocsis2006_2}, using a selection strategy inspired by the UCB1 algorithm \citep{Auer2002}. At each decision point, the selected child is given by
\begin{equation}\label{UCT}
    \mathop{\mathrm{argmax}}_{v' \in \text{\Call{Children}{v}}} \left( \frac{Q(v')}{N(v')} + C_p \sqrt{\frac{2\ln N(v)}{N(v')}} \right),
\end{equation}
\noindent
where $\textnormal{\textsc{Children}{(v)}}$ returns the set of children of node $v$ and $C_p$ is a chosen exploration constant. $N(v)$, the number of visits to node $v$, and $Q(v)$, the total return obtained when visiting node $v$, are updated during the backpropagation phase (cf. Alg. \ref{alg:mcts_backprop}).

\subsection{Actions as independent sets}\label{sec:mwis_of_actions}
In the traditional MCTS algorithm, a child node is always obtained after applying only one of the available actions from the parent node. In \textsc{AtomTreeSearch}, we consider a subset of the available actions which satisfies the following properties: (i) all actions in this set can be collectively performed in any order to reach the same state and (ii) the reward received for each action in this set does not depend on the order in which the actions are performed. We can view such a subset of independent actions as a single collective action. Therefore, we modify the MCTS algorithm as follows (cf. Alg. \ref{alg:qmcts_select} and Alg. \ref{alg:qmcts_expand}): actions taken in \textsc{AtomTreeSearch} become independent sets of individual actions, and a node is considered fully expanded if all independent sets of actions were used to obtain a child. As this process may generate many independent sets, and therefore many collective actions, we keep only a certain number of sets for which the sums of the rewards are the greatest. The rest of the algorithm remains the same.

Note that in Alg. \ref{alg:qmcts_expand}, the state $s'$ is obtained by sampling from the distribution $p\left(\cdot\middle| s,a\right)$, the state transition function (the algorithm assumes a stochastic state transition function to be general; we would have $s^\prime=p\left(s,a\right)$ in the deterministic case). In other words, when starting in state $s$ and performing action $a$, a transition to state $s^\prime$ happens with probability $p\left(s^\prime\middle| s,a\right)$.

Obtaining the independent sets of actions is done using Alg. \ref{alg:actions_mwis} which we now describe. A conflict graph $G$ is built where each action available from an MDP state is mapped to a vertex, the weight of the vertex being the expected reward received when performing this individual action from this state (represented as $r(s,\mathcal{A}_i)$ in Alg. \ref{alg:actions_mwis}). An edge connects two vertices if the associated actions can't be part of a same set satisfying properties (i) and (ii), otherwise no edge connects the vertices. A quantum subroutine is then used to provide many candidate solutions to the MWIS problem defined on $G$ (see line 11 of Alg. \ref{alg:actions_mwis}). The independent sets of action, i.e., the collective actions that can be taken in \textsc{AtomTreeSearch}, are then retrieved by mapping each vertex of the obtained solution sets to their corresponding action. Note that executing Alg. \ref{alg:actions_mwis} requires a call to a quantum computer only the first time the algorithm is invoked for the state associated with a given node. Subsequent invocations of $\textnormal{\textsc{MwisOfActions}}$ for the same node incur negligible computational cost as the previously computed set is retrieved from memory. In Alg. \ref{alg:qmcts_select}, $\textnormal{\textsc{MwisOfActions}}$ is employed to determine whether the node currently considered is fully expanded. In contrast, Alg. \ref{alg:qmcts_expand} calls $\textnormal{\textsc{MwisOfActions}}$ to identify independent sets that have not yet been used and can therefore be utilized to generate new child nodes.

\begin{algorithm}[t]
\caption{\textnormal{\textsc{Select}}($v$)}\label{alg:qmcts_select}
\textbf{Input:} The root node $v$\\
\textbf{Output:} The node selected to be expanded
\begin{algorithmic}[1]
\State $\mathcal{W} \gets$ \Call{MwisOfActions}{$s(v), nb\_samples$}
\State $is\_fully\_expanded \gets |\mathcal{W}|==|$\Call{Children}{$v$}$|$
\While{$\neg$ \Call{terminal}{$s(v)$}$\ \land \ is\_fully\_expanded$}
    \vspace{1sp}
    \State $v \gets \displaystyle\mathop{\mathrm{argmax}}_{v' \in \text{\Call{Children}{v}}} \left( \frac{Q(v')}{N(v')} + C_p \sqrt{\frac{2\ln N(v)}{N(v')}} \right)$
    \vspace{2mm}
    \State $\mathcal{W} \gets$ \Call{MwisOfActions}{$s(v), nb\_samples$}
    \State $is\_fully\_expanded \gets |\mathcal{W}|==|$\Call{Children}{$v$}$|$
\EndWhile
\State \textbf{return} $v$
\end{algorithmic}
\end{algorithm}

\begin{algorithm}[t]
\caption{\textnormal{\textsc{Expand}}($v$)}\label{alg:qmcts_expand}
\textbf{Input:} The node $v$ to be expanded\\
\textbf{Output:} The new child node
\begin{algorithmic}[1]
\State $s \gets s(v)$
\State $\mathcal{W} \gets$ \Call{MwisOfActions}{$s, nb\_samples$}
\State choose $\mathcal{M} \in$ untried action sets from $\mathcal{W}$
\For{$a \in \mathcal{M}$}
    \State $s' \sim p(\cdot | s,a)$
    \State $s \gets s'$
\EndFor
\State create node $v'$ from $s$
\State add new child $v'$ to $v$
\State \textbf{return} $v'$
\end{algorithmic}
\end{algorithm}

\begin{algorithm}[t!]
\caption{\textsc{Backpropagate}($v,R$)}\label{alg:mcts_backprop}
\begin{tabular}[t]{@{}l@{\ }l@{}}
\textbf{Input:} &Node $v$,\\
                &Return $R$ from a simulation starting at $v$
\end{tabular}
\begin{algorithmic}[1]
\While{$v$ is not null}
    \State $N(v) \gets N(v) + 1$\;
    \State $Q(v) \gets Q(v) + R$\;
    \State $v \gets \text{parent of } v$\;
\EndWhile
\end{algorithmic}
\end{algorithm}

\subsection{Quantum approach to the MWIS subroutine}\label{sec:naqc}

The quantum subroutine used on line 11 of Alg. \ref{alg:actions_mwis} is described in Alg. \ref{alg:sample_mwis} and natively runs on a neutral-atom quantum computer, although it could be adapted to run on other quantum computing hardware modalities as discussed at the end of this section. Neutral-atom quantum computing is done using an array of alkali metal atoms, typically rubidium \citep{Henriet2020, Leclerc_2024}. A time-dependent laser characterized by the amplitude $E(t)$, frequency $\omega(t)$ and phase $\phi(t)$ of its electric field is used to drive transitions between electronic states of the valence electron. A pair of electronic levels then forms a qubit. Here, we use the ground state of the atom $\ket{g}$  as the logical $\ket{0}$ and the Rydberg state $\ket{r}$, a highly excited electronic state, as the logical $\ket{1}$. The readout is performed using fluorescence imaging: atoms in $\ket{0}$ are detected by fluorescence, whereas atoms in $\ket{1}$ are detected by absence of fluorescence.

\begin{algorithm}[t]
\caption{\textnormal{\textsc{MwisOfActions}}($s, nb\_samples$)}\label{alg:actions_mwis}
\begin{tabular}[t]{@{}l@{\ }l@{}}
\textbf{Input:} &An MDP state $s$,\\
                &A number of samples $nb\_samples$\\
\end{tabular}\\
\textbf{Output:} A set of maximal independent sets of actions
\begin{algorithmic}[1]
\If{\Call{MwisOfActions}{} was previously called with $s$}
    \State \textbf{return} the saved set
    \Comment{See line 13}
\EndIf
\State \begin{tabular}[t]{@{}l@{\ }l@{}}
$\mathcal{A} \gets$ & the set of all individual actions available \\
                    & from $s$
\end{tabular}
\State $G \gets$ new empty graph
\For{$i=1, \dots,|\mathcal{A}|$}
    \State add vertex $i$ to $G$ with weight $w_i=r(s,\mathcal{A}_i)$
\EndFor
\For{$i=1, \dots,|\mathcal{A}|$}
    \For{$j=i+1, \dots,|\mathcal{A}|$}
        \State \begin{tabular}[t]{@{}l@{\ }l@{}}
        &\textbf{if} $\mathcal{A}_i$ and $\mathcal{A}_j$ can't form a set satisfying\\
        & conditions (i) and (ii) \textbf{then}
        \end{tabular}
            \State \hspace*{\algorithmicindent} add edge ($i, j$) to $G$
    \EndFor
\EndFor
\State $\mathcal{F} \gets$\Call{SampleMwisSolutions}{$G, nb\_samples$}
\State $\mathcal{W} \gets \{\{\mathcal{A}_i|i\in \mathcal{D}\}|\mathcal{D} \in \mathcal{F}\}$
\State Save $\mathcal{W}$ in case \Call{MwisOfActions}{} is called with $s$ again later
\State \textbf{return} $\mathcal{W}$
\end{algorithmic}
\end{algorithm}

Atoms can be arranged in an arbitrary configuration in two or three dimensions by creating a layout of optical traps. Typically, a global laser shining on all atoms is used with a Rabi frequency $\Omega(t) \propto E(t)$ and a detuning $\delta(t) = \omega(t) - (E_r - E_g)/\hbar$ where $E_g$ and $E_r$ are the energies associated to $\ket{g}$ and $\ket{r}$ respectively. The detuning of this laser can be made to address on each atom locally using a Detuning Map Modulator (DMM). The effect of such a DMM is equivalent to applying a pulse with zero Rabi frequency and a negative detuning $\Delta(t) $ modulated by a weight $\epsilon_i \in [0, 1]$ specific to each atom. In the presence of a DMM, the Hamiltonian of a system of $N$ Rydberg atoms coupled to a laser is given by:
\begin{equation}\label{qpu_hamiltonian}
\begin{split}
    H(t) = &\frac{\hbar\Omega(t)}{2} \sum_{i=1}^{N} \hat{\sigma}_i^x - \hbar\delta(t) \sum_{i=1}^{N} \hat{n}_i\\
    &- \hbar\Delta(t) \sum_{i=1}^{N} \epsilon_i\hat{n}_i + \sum_{i=1}^{N}\sum_{j=i+1}^{N} \frac{C_6}{R_{ij}^6}\hat{n}_i\hat{n}_j,
\end{split}
\end{equation}

\noindent
where we assumed $\phi(t)=0$ to simplify Eq. \ref{qpu_hamiltonian}. The Pauli $\hat{\sigma}_x$ operator for atom $i$ is given by ${\hat{\sigma}_i^x=({\ket{r}\bra{g}}_i + {\ket{g}\bra{r}}_i)/\sqrt{2}}$ and $\hat{n}_i = \ket{r}\bra{r}_i$ is the projector on the Rydberg state for atom $i$ so that $\hat{n}_i\ket{r}_i = \ket{r}_i$ and $\hat{n}_i\ket{g}_i = 0$. The distance between atom $i$ and atom $j$ is given by $R_{ij}$ and $C_6$ is a coefficient that depends on the chosen principal quantum number of the Rydberg state. The final term, arising from van der Waals interactions, induces a significant energy shift of the doubly excited state $\ket{r}_i\ket{r}_j$ when $R_{ij} \leq R_b$, where the Rydberg blockade radius $R_b$ is given by
\begin{equation}\label{blockade_radius}
R_b = \left(\frac{C_6}{\hbar\sqrt{\Omega^2 + \delta^2}}\right)^\frac{1}{6}.
\end{equation}

In this regime, the Rydberg blockade mechanism prevents the simultaneous excitation of both atoms: a laser resonant with the single-atom transition can no longer excite both atoms simultaneously.

Coming back to the resolution of the MWIS problem, we formulate $MWIS(G)$ as a Quadratic Unconstrained Binary Optimization (QUBO) problem given by
\begin{equation}\label{mwis_qubo}
    \mathop{\mathrm{argmin}}_{\textbf{x} \in \{0, 1\}^{|V|}} \left( -\sum_{u \in V} w_u x_u + \alpha \sum_{\{u,v\} \in E} x_u x_v \right),
\end{equation}
\noindent
where $G(V,E)$ is the graph with vertex set $V$ and edge set $E$ on which the MWIS problem is to be solved, $\textbf{x} = \{x_1, ..., x_{|V|}\}$ is a binary vector, with $x_u=1$ if vertex $u$ is included in the solution, and $x_u=0$ otherwise. The penalty coefficient $\alpha > 0$ ensures that the solution is a valid independent set. In practice one requires $\alpha > \max_{uv} \min(w_u, w_v)$.

The method proposed in Alg. \ref{alg:sample_mwis} is a quantum annealing \citep{Yarkoni_2022} protocol that runs on a neutral-atom quantum computer aiming to return high quality solutions to this QUBO problem. Similarly to the adiabatic quantum algorithm, quantum annealing involves preparing a system in an initial Hamiltonian $H_0$ and slowly evolving it towards a problem Hamiltonian $H_P$ whose low-energy states we aim to sample from. Thus, the QUBO resolution requires formulating the problem Hamiltonian $H_P$ in the form of Eq. \ref{qpu_hamiltonian}, such that its ground state encodes the solution to the QUBO formulation in Eq. \ref{mwis_qubo}. Each binary variable $x_u$ is mapped to a qubit. Since the interaction term in Eq.~\ref{qpu_hamiltonian} depends only on interatomic distances, the Rydberg blockade mechanism is used to enforce the independent set constraint. For instance, if atoms $a_u$ and $a_v$, corresponding to vertices $u$ and $v$, are separated by less than the Rydberg blockade radius, an effective penalty $\alpha=\infty$ is imposed as both atoms cannot be simultaneously excited. To encode the MWIS problem for a general graph, it is therefore necessary to place atoms at prescribed coordinates in a two-dimensional plane such that, for every edge in the graph, the atoms associated with its ends lie within the Rydberg blockade radius of one another. Graphs that admit such a representation are known as unit-disk graphs, and determining the corresponding atomic positions constitutes the embedding problem. Embedding a graph as a unit-disk graph in two dimensions is NP-hard and, for arbitrary random graphs, a feasible embedding may not exist \citep{Breu1998}.
\begin{algorithm}[t]
\caption{\textnormal{\\ \textsc{SampleMwisSolutions}}($G,nb\_samples$)}\label{alg:sample_mwis}
\begin{tabular}[t]{@{}l@{\ }l@{}}
\textbf{Input:} &A graph $G(V,E)$,\\
                &A number of samples $nb\_samples$\\
\end{tabular}\\
\begin{tabular}[t]{@{}l@{\ }l@{}}
\textbf{Output:} &A set of maximal weighted independent sets\\
                 &of $G$\\
\end{tabular}
\begin{algorithmic}[1]
\State Embed $G(V,E)$ into a unit disk representation $G'(V',E')$
\State Determine DMM weights using vertex weights
\State $\mathcal{F} \gets \emptyset$
\For{$i=1, \dots,nb\_samples$}
    \For{$v' \in V'$}
        \State \begin{tabular}[t]{@{}l@{\ }l@{}}
        &Place an atom in the optical tweezer array\\
        &at the position of $v'$
        \end{tabular}
    \EndFor
    \State Initialize the system in state $\ket{\psi_0}=\ket{0}^{\otimes|V|}$
    \State Apply the laser pulse sequence to obtain $\ket{\psi_f}$
    \State $b \gets \Call{Measure}{\ket{\psi_f}}$
    \State $b \gets \Call{VertexReduction}{G, b}$
    \State $b \gets \Call{VertexAddition}{G, b}$
    \State \begin{tabular}[t]{@{}l@{\ }l@{}}
    $\mathcal{D} \gets$ & the maximal weighted independent set of\\
                        & $G$ represented by $b$
    \end{tabular}
    \State $\mathcal{F} \gets \mathcal{F} \cup \{\mathcal{D}\}$
\EndFor
\State \textbf{return} $\mathcal{F}$
\end{algorithmic}
\end{algorithm}

To overcome this difficulty, various heuristic strategies have been proposed in previous works \citep{coelho2022, cazals2025_2, Schuetz2025, perron2026leveraging}, one of which relies on the force-directed Fruchterman–Reingold algorithm. This approach does not guarantee a valid unit-disk embedding, leading to discrepancies between $G$ and its unit-disk representation: spurious edges may appear, and required edges may be missing. In such cases, $H_P$ is no longer a faithful representation of the intended QUBO problem, and its ground state may not correspond to a valid MWIS solution. Efficient post-processing techniques, based on vertex reduction and addition as described by \cite{ebadi2022}, or other such techniques (see for example \cite{Leclerc_2025}) can be used to recover valid solutions. The use of both techniques ensures that the recovered independent set is always a maximal independent set.

Once the embedding is obtained, the qubits are initialized in their individual ground state $\ket{\psi_0}=\ket{0}^{\otimes |V|}$, and the initial Hamiltonian $H_0$ is evolved towards $H_P$ by continuously modulating the laser field. The duration of this pulse is of time $T$, which is taken to be smaller than the decoherence time of the quantum system beyond which environmental noise overcomes the proposed protocol. The initial detuning is chosen to be strongly negative and the Rabi frequency is null so $\ket{\psi_0}$ is the ground state of $H_0$. The final detuning at time $t=T$ is set to encode the vertex weights $w_u$ according to
\begin{equation}\label{weights_encoding}
    cw_u = \delta(T) + \epsilon_u \Delta(T),
\end{equation}
\noindent
where $c$ is a normalization constant accounting for laser limitations, and $\delta$, $\epsilon_u$ and $\Delta$ are defined as in Eq.~\ref{qpu_hamiltonian}.

Because the total pulse duration is limited, the adiabatic condition is generally not satisfied. Therefore, quantum annealing is performed instead of adiabatic quantum computing and can be understood as a heuristic quantum algorithm. As a result, the system may accumulate significant amplitude in low-lying excited states during the evolution. The final state $\ket{\psi_f}$, obtained at the end of the annealing, is therefore a superposition of candidate MWIS solutions. This property can be exploited by executing quantum annealing multiple times to sample a diverse set of candidate solutions. Each measurement of $\ket{\psi_f}$ collapses the wave function, yielding a bitstring where the bit at position $u$ is 1 if vertex $u$ is selected in the solution and 0 otherwise. Finally, post-processing techniques are applied to these bitstrings individually.

Different quantum algorithms designed to solve combinatorial optimization problems could be used to solve the QUBO problem such as QAOA. The algorithm solving the QUBO problem could also be executed on different quantum computing architectures such as those based on superconducting qubits \citep{Pelofske_2023a}, trapped ions \citep{Pelofske_2023b} or photonic qubits \citep{cazalis2024}, though here we explicitly use the strong Rydberg blockade constraints present in neutral atom platforms to encode the constrained combinatorial structure to sample from. As such, \textsc{AtomTreeSearch} leverages the strong atomic interactions to encode the independence between individual actions, and quantum protocols enable a diverse yet high-quality sampling of the resulting collective actions.

\section{Application to the TSP}\label{sec:TSP_app}

To evaluate the performance of our algorithm, we apply it to the Traveling Salesman Problem (TSP). This section first introduces the TSP and reviews the main approaches used to solve it in Section \ref{sec:TSP}. We then explain how \textsc{AtomTreeSearch} can be adapted to this problem (Section \ref{sec:method}) and describe the datasets (Section \ref{sec:dataset}) and classical solvers employed for benchmarking (Section \ref{sec:solver}).

\subsection{The Traveling Salesman Problem}\label{sec:TSP}

The TSP can be stated as follows: given a graph with weighted edges, find a cycle visiting every vertex once and only once such that the length of the cycle (calculated as the sum of the traversed edge’s weight) is minimized. In the TSP, vertices are also referred to as cities and Hamiltonian cycles are also referred to as tours.

Certain methods from operations research, notably linear programming (e.g., branch-and-cut algorithms implemented in solvers such as Concorde \citep{applegate2003} and Gurobi \citep{gurobi}) and dynamic programming, most prominently the Held–Karp algorithm \citep{held1962}, can solve moderate-sized instances of this problem exactly. However, their computational time grows exponentially with the number of cities.

Constructive heuristics such as the Nearest Neighbor \citep{rosenkrantz1977}, Greedy \citep{bentley1992}, Nearest Insertion \citep{johnson2002}, Farthest Insertion \citep{johnson2002}, Clarke–Wright Savings \citep{clarke1964}, and Christofides \citep{christofides1976} algorithms, as well as improvement heuristics such as the 2-Opt \citep{flood1956, croes1958}, 3-Opt \citep{lin1965} and Lin–Kernighan \citep{lin1973} algorithms, and metaheuristics such as Tabu Search \citep{glover1986}, Simulated Annealing \citep{Kirkpatrick1983}, Genetic Algorithms \citep{Brady1985} and Ant Colony Optimization \citep{Dorigo1997} can instead deliver good-quality solutions within reasonable time. Nevertheless, these approaches often rely on structural assumptions or empirical parameter tuning. For instance, Christofides’ algorithm applies only to metric instances of the TSP. \cite{johnson2002} present an extensive analysis and discussion of the performance achieved by several of these approaches.

More recently, machine learning–based methods have demonstrated the ability to compete with traditional heuristics on certain instance sets. Improvement-based approaches (e.g., Learned 2-Opt \citep{Costa2020}, Neural 3-Opt \citep{Sui2021}), which leverage operations research heuristics, tend to yield better results than constructive approaches (e.g., Ptr-Net \citep{Vinyals2015}, AM \citep{kool2019}, S2V-DQN \citep{dai2017}), though typically at the cost of longer runtimes. Regarding MCTS, it has been extensively studied as a methodology to tackle the TSP and vehicle routing problems (VRPs) in general \citep{Powley2012, Perez2012, Mandziuk2015, Mandziuk2017, Nguyen2020, xing2020, Fu2021, qiu2022, sun2023, Barletta2023, pan2024}. These methods generally require a training phase, whereas \textsc{AtomTreeSearch} does not. When applied to the TSP, the works of \cite{Fu2021, sun2023} and \cite{pan2024} report that competitive results can be obtained within reasonable computational time, with optimality gaps below 1\% on TSPLIB instances comprising up to 100 cities. Still, machine learning–based methods remain limited in their ability to generalize to larger problem instances and do not consistently outperform established heuristics.

The TSP can be formulated as an Ising model \citep{Lucas2014}, solvable via quantum annealing or adiabatic quantum computation \citep{martonak2004, heim2017, Kieu2019} using $(N-1)^2$ qubits, where $N$ is the number of vertices (cities). Using a similar approach, the Variational Quantum Eigensolver (VQE) algorithm and QAOA \citep{Bartschi2020,ruan2020, Qian2023, bourreau2023, Khumalo2025} have also been used to tackle the TSP. However, these methods often require qubit counts, coherence times and hardware connectivity exceeding what current hardware offers. Alternatively, \cite{srinivasan2018} use quantum phase estimation and amplitude amplification to achieve a quadratic speedup over a classical brute-force search yielding a $O(\sqrt{(N-1)!/2})$ time algorithm. Hybrid frameworks have also been developed, such as a quantum tabu algorithm \citep{osaba2021}, a quantum ant colony optimization algorithm \citep{qiu2024} and an algorithm using path slicing methods to divide the TSP into smaller subproblems solved by a quantum computer \citep{liu2024}. Finally, quantum computing has been integrated with machine learning methods in other works. Examples include applying quantum Q-learning \citep{Xu2024} and quantum support vector machines \citep{Mohanty2024} to the generalized vehicle routing problem. For an overview of quantum computing methods applied to routing problems, see \cite{Osaba2022} and \cite{phillipson2025}.
\vfill

\subsection{Method}\label{sec:method}
Our method first requires an initial solution to the TSP. We use one of the constructive heuristics mentioned above to quickly return one. This initial tour is then considered as an MDP state and is given as the root node of our modified MCTS algorithm. Thus, the initial state provided to Alg. \ref{alg:mcts_search} is the initial solution to the TSP.

\begin{figure}[t]
  \centering
  \includegraphics[width=0.5\textwidth]{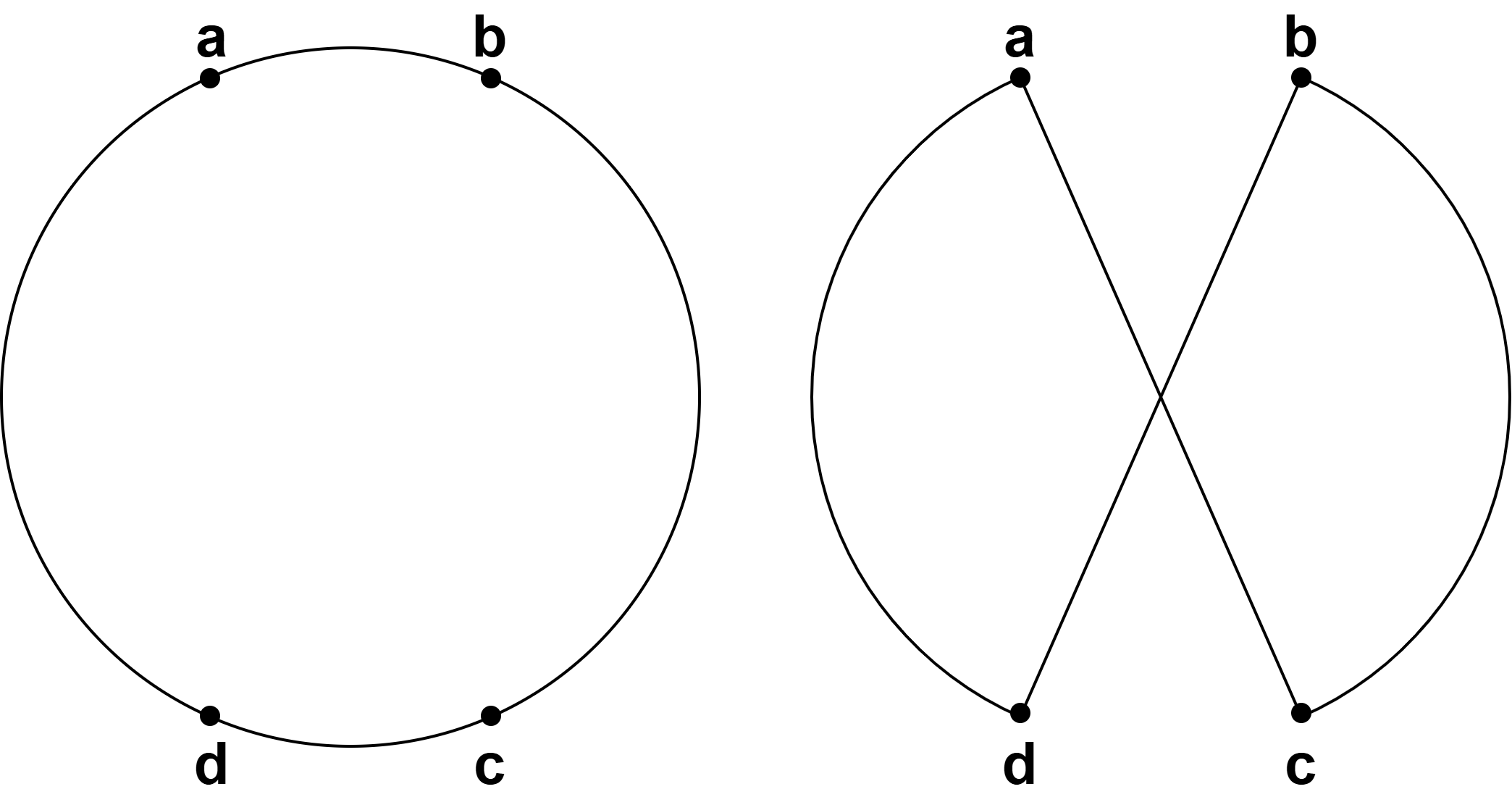}
  \caption{Example of 2-Opt operation. Edges (a,c) and (b,d) replace edges (a,b) and (c,d).}
  \label{fig:2-opt}
\end{figure}

Algorithms \ref{alg:mcts_search} - \ref{alg:sample_mwis} are implemented. We define an individual action as a 2-Opt operation in which two edges of a tour are removed and the remaining segments are reconnected in reversed order (see Fig. \ref{fig:2-opt}). Two individual actions are considered independent if their corresponding 2-Opt operations do not share a removed edge or an inserted edge. The weight of an action is given by the reduction in the cycle length provided by the associated 2-Opt operation. An operation reducing the cycle length has a positive weight.

\begin{figure*}[t]
  \centering
  \includegraphics[width=1\textwidth]{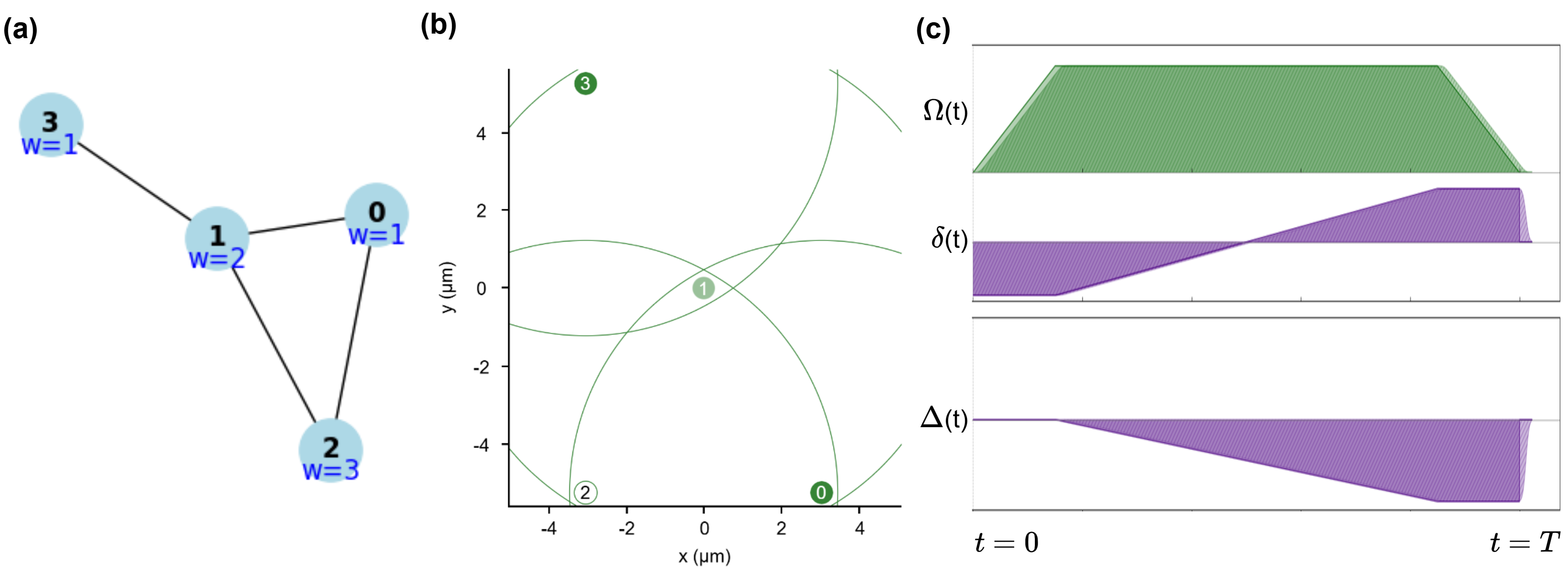}
  \caption{(a) Conflict graph $G$ of actions with their associated weights. (b) Embedding of $G$ on a triangular lattice. Atoms are shown in shades of green, with the color intensity proportional to their DMM weights. The green circle surrounding each atom represents its Rydberg blockade radius. (c) The solver pulse. $\Omega(t)$ ramps linearly from 0 to $\Omega_{max}$ over $t \in [0, 0.15T]$, with a global negative detuning held constant. During the sweep phase ($t \in [0.15T, 0.85T]$), $\Omega$ remains fixed while the detuning evolves linearly. In the final phase, $\Omega$ linearly decreases to 0 while the detuning is fixed. The final detuning is chosen to encode the vertex weights $w_u$ through the relation given in Eq. \ref{weights_encoding}.}
  \label{fig:pulse_seq}
\end{figure*}

To realize the embedding step in Alg. \ref{alg:sample_mwis}, a heuristic embedding strategy based on the Fruchterman-Reingold force-directed algorithm and described in detail by \cite{perron2026leveraging} is implemented. Other methods to encode the relevant MWIS problem in a spatial arrangement of atoms could be used. Then, we choose to test a simple quantum annealing protocol represented by the pulse waveform shown in Fig. \ref{fig:pulse_seq}. The strength of both $\Omega_{max}$ and $\delta_{max}=-\delta_{min}$ are set by the blockade radius needed to best embed the conflict graph in the spatial arrangement of atoms. In this proof of concept, the maximum duration of the quantum evolution is 5 µs, and the quantum evolution is emulated using classical simulation tools such as the \texttt{QuTiP} \citep{johansson2012, johansson2013, qutip5}, \texttt{EMU-SV} \citep{bidzhiev2025} and \texttt{EMU-MPS} \citep{bidzhiev2025} libraries, depending on the size of the quantum system to be emulated. Vertex reduction and addition techniques as described by \cite{ebadi2022,cazals2025_1} and \cite{perron2026leveraging}, are implemented to post-process the sampled bitstrings into valid MWIS solutions. Other post-processing techniques exist, such as those proposed by \cite{Leclerc_2025}, and could be used.

In practice, quantum computers only have $M$ available qubits, while there is in the order of $O(N^2)$ possible 2-Opt operations (vertices) in the conflict graph $G$. This limitation means that the graph generated in Alg. \ref{alg:actions_mwis} sometimes can’t be passed to Alg. \ref{alg:sample_mwis} because this algorithm requires as many qubits as there are vertices in this graph. To circumvent this issue, the graph is partitioned into subgraphs having a number of vertices smaller or equal to $M$. Three methods are implemented  to achieve this: (i) by repeatedly calling the Kernighan-Lin partitioning algorithm until the largest subgraph created by this partitioning is smaller than $M$ (ii) by generating subgraphs that are guaranteed to be disconnected using the method described in the next paragraph (iii) by randomly selecting connected vertices until the generated subgraph is of the desired size. We respectively call these methods the ``KL", ``disconnected" and ``random" partitioning methods.

Method (ii) begins by marking all vertices as unvisited. While unvisited vertices remain, it iteratively performs the following procedure. First, an unvisited vertex is selected at random to initialize a new subgraph and is marked as visited. Starting from this vertex, the method performs a randomized breadth-first expansion: the vertex is added to the current subgraph and its unvisited neighboring vertices are marked as visited and randomly shuffled before being inserted into a queue from which subsequent vertices are selected for inclusion in the subgraph. This expansion process continues until either the queue becomes empty or a predefined maximum subgraph size is reached. Since every neighbor of a vertex included in the subgraph is immediately marked as visited, no such vertex can later belong to a different subgraph. Consequently, the generated subgraphs are pairwise disconnected.

Solutions to the MWIS problem for each of the subgraphs generated by one of the methods (i) - (iii) are then obtained using Alg. \ref{alg:sample_mwis}. For methods (i) and (ii), if $k$ different sets are returned by Alg. \ref{alg:sample_mwis} for $n$ subgraphs, then there are $k^n$ different ways to combine these sets. To limit the number of proposed solutions to the original conflict graph, we randomly combine the subgraph solutions until $f(k,n)$ solutions are generated, where $f(k,n)$ is given by
\[
f(k,n) =
\begin{cases}
k^n, & \text{if } k^n \le 125, \\
k n, & \text{if } k^n > 125 .
\end{cases}
\]
{\setlength{\parskip}{4pt}
\noindent
Finally, vertex reduction and addition is performed on these solutions to ensure validity with respect to the MWIS problem. The $k$ best independent sets are returned by Alg. \ref{alg:actions_mwis}.

During the expansion phase (Alg. \ref{alg:qmcts_expand}), the 2-Opt operations that are part of the same independent set are applied one after another to the tour associated to the current state. Therefore, in this implementation the state transition function is deterministic (line 5 of Alg. \ref{alg:qmcts_expand} reads $s' \leftarrow p(s,a)$ for this implementation). The state $s'$ on line 5 of Alg. \ref{alg:qmcts_expand} represents the new temporary TSP solution obtained after performing a given 2-Opt operation to the solution associated to state $s$. Since we do not check at this point that all vertices in the performed 2-Opt operation come after or before the vertices in all other 2-Opt operations performed during the expansion phase, it might happen that subtours are created. Thus, as the last step of the expansion phase, we stitch the subtours together if needed and therefore reobtain a Hamiltonian cycle. We use the heuristic based on a minimum spanning tree described by \cite{Kahng2004}. Only the exact patching algorithm described in this paper is used, though the alternating patching algorithm could also be used in future implementations. When stitching fails (this might happen with graphs that are not complete), the expansion phase simply consists of performing the best 2-Opt operation found. The cycle obtained from this procedure is associated to a state which will correspond to the new child node.

The simulation phase ($\textnormal{\textsc{Simulate}}$ in Alg. \ref{alg:mcts_search}) is done using a simulated annealing algorithm. Simulated annealing was first introduced by \cite{Kirkpatrick1983} and has since been extensively studied in the context of the TSP \citep{Aarts1988, Allwright1989, Jeong1991, Tian1993, Geng2011}. In the implementation considered here, which closely follows the approach proposed by \cite{Kirkpatrick1983}, a random 2-Opt operation is selected and is performed if the length of the new cycle is shorter. Otherwise, the increase of the cycle length is used as the energy $E$ in the Metropolis acceptance rule: the worsening operation is performed if a random number $p \in U(0,1)$, where $U$ is the uniform distribution, is such that $p<e^{-E/T}$, with $T$ the temperature. We use an initial temperature of 100 and a cooling rate of 0.98. This is repeated for 20 iterations, therefore simulating the serial application of 2-Opts on the state $s$.

The length of the cycle obtained from the simulation phase is used to compute the return $R$ which appears in the backpropagation phase (Alg. \ref{alg:mcts_backprop}). In the general setting of a minimization problem, this could be stated as
\begin{equation}\label{tsp_sim_return}
    R=\frac{UB-C(s)}{UB - LB},
\end{equation}
where $UB$ is the upper bound for the value of a solution, $LB$ is a lower bound and $C(s)$ is the cost of the current solution. In the case of the TSP, we take the maximum cycle length ($UB$) and minimum cycle weight ($LB$) to be given by
\begin{equation}\label{tsp_ub_lb}
\begin{split}
    UB&=\sum_{e \in MST(G)} w(e) + \max_{e \in E} w(e)\\
    LB&=\sum_{e \in mST(G)} w(e) + \min_{e \in E} w(e),
\end{split}
\end{equation}
where $MST(G)$ and $mST(G)$ are respectively the set of edges in the maximum spanning tree and the set of edges in the minimum spanning tree of graph $G$, $E$ is the set of edges in graph $G$ and $w(e)$ is the weight associated to edge $e$. These are two simple bounds and other more refined bounds could be used. For implementations of \textsc{AtomTreeSearch} to other CO problems, other bounds can be used. The length of the final cycle obtained from the simulation phase is used as $C(s)$. This gives $R\in\left[0,1\right]$, however we find that due to the simple $UB$ and $LB$ we chose, the value of $R$ falls in a much smaller range of values during the execution of the algorithm. In order to prevent Eq. \ref{UCT} from being dominated by the exploration term, which is the case when the range of $R$ is small, we empirically choose the exploration constant in Alg. \ref{alg:qmcts_select} to be $C_p=0.02$ which effectively mitigates this effect.

We finally note that the algorithm could be stopped after any number of iterations and provide multiple valid tours (e.g., the tours corresponding to all nodes or the best tours found during the search in any phase of Alg. \ref{alg:mcts_search}). Since our method relies on performing 2-Opt operations, it is also worth noting that it can tackle TSP instances that are not metric, which is a challenging setting for other heuristics such as Christofides' algorithm.

\subsection{Dataset}\label{sec:dataset}

We use two sources of TSP instances. The first is the well-known TSPLIB \citep{reinelt1991}, and the second is a custom dataset of TSP instances obtained from a random Euclidean graph generator. In the first case, we use all instances with 100 cities or less.

In the latter case, the graph generator takes as input the number of vertices $N$ and a density parameter $d \in [0, 1]$. A set of $N$ random points, corresponding to the vertex coordinates, is generated within the square $[0, 100]^2$. The vertices are numbered from $1$ to $N$, and an edge is added between each vertex $u$ and $(u+1) \bmod N$, thereby forming an initial cycle. The graph is then guaranteed to have a Hamiltonian cycle. Additional random edges are then added until the graph contains $d \, \frac{N(N-1)}{2}$ edges. The weight of each edge is given by the Euclidean distance between its two endpoints. The generator returns the graph and the initial cycle.
}

\subsection{Comparative solvers}\label{sec:solver}
To quantify the interest of using a quantum computer in our method, we also run our algorithm using different classical MWIS solution samplers in Alg. \ref{alg:actions_mwis}. We first employ a random sampler that constructs an independent set by iteratively selecting a random vertex, adding it to the independent set, and removing that vertex along with its neighbors from further consideration. This could effectively be described as a random maximal independent set sampler. We also use three classical sampling methods described by \cite{perron2026leveraging}. These methods are based respectively on greedy selection, simulated annealing, and integer linear programming with diversification (ILP+DIV). The latter returns the best $k$ independent sets, but has exponential runtime. It acts as a baseline. We name our alternate \textsc{AtomTreeSearch} algorithms as \textsc{RdmTreeSearch}, \textsc{GreedyTreeSearch}, \textsc{SATreeSearch} and \textsc{ILPTreeSearch} when we use respectively the random sampler, greedy sampler, simulated annealing sampler and integer linear programming sampler to obtain solutions to the MWIS problem defined in Alg. \ref{alg:actions_mwis}.

We also compare the performance of our method with the solutions obtained by classical TSP solvers, namely a simulated annealing algorithm performing 2-Opt operations and the \texttt{OR-Tools} software suite \citep{ortools}. The simulated annealing algorithm used is the same as the one used in the simulation phase of the tree search described earlier, but takes different parameter values. The initial temperature is set to 100, the cooling rate is 0.999 and the algorithm performs $10\ 000$ iterations which is far more than the number of iterations used in the simulation phase. When using \texttt{OR-Tools}, we first obtain an initial solution with the \texttt{PATH\_CHEAPEST\_ARC} option and subsequently set the local search option to \texttt{AUTOMATIC}, allowing the solver to select an appropriate local search strategy. Optimal solutions obtained using the Concorde solver \citep{applegate2003} are used to compute the optimality gaps of solutions generated by the other solvers.

All five tree search algorithms and the simulated annealing algorithm use the initial cycle provided by the random Euclidean graph generator as a starting point when solving a random TSP instance. When solving a TSPLIB instance, these solvers all use a same cycle obtained using the greedy construction heuristic implemented in the \texttt{NetworkX} library \citep{Hagberg2008} and described by \cite{bentley1992}.
\vfill

\section{Results}\label{sec:results}

\begin{figure*}[t]
  \centering
  \includegraphics[width=\textwidth]{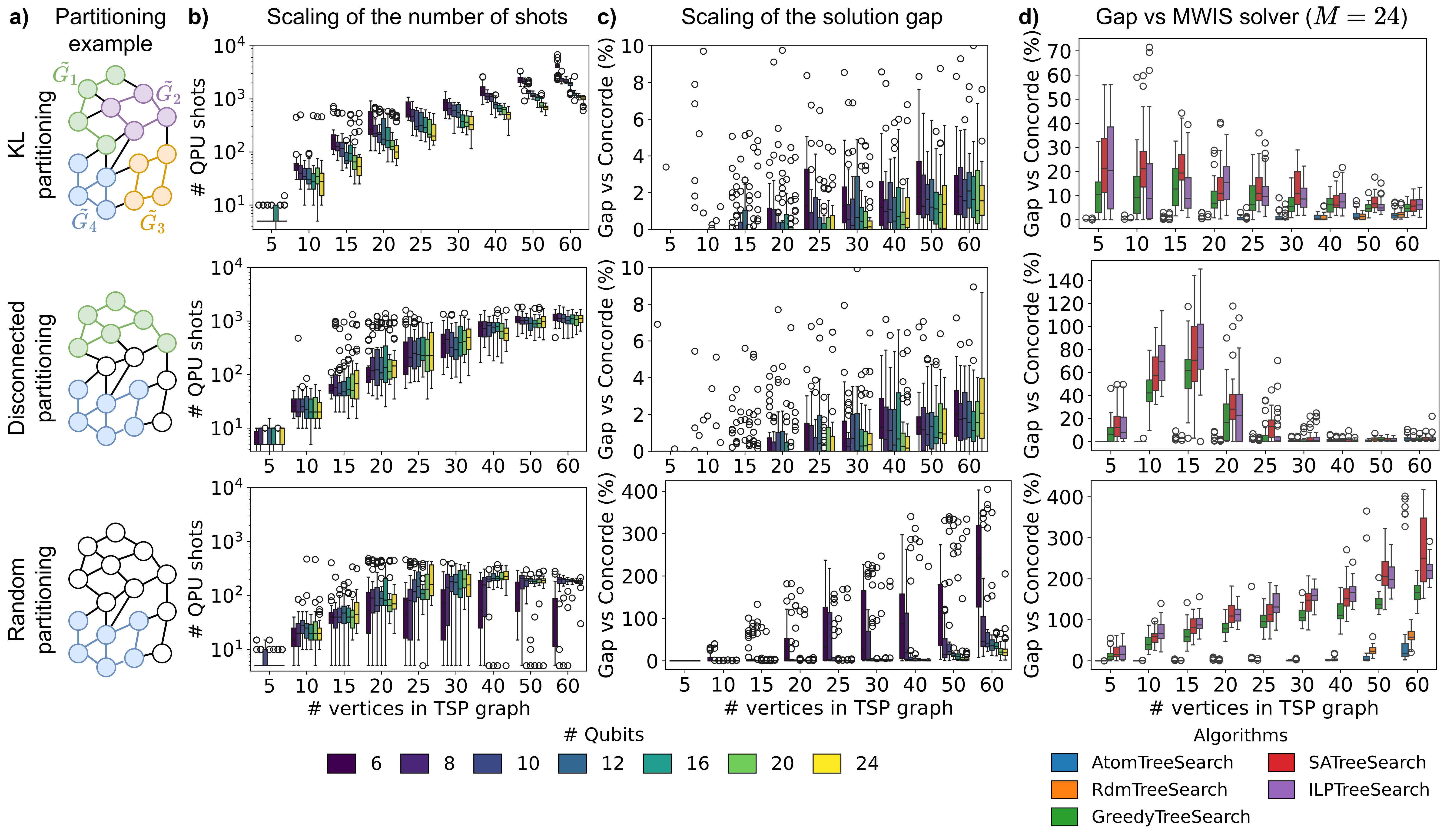}
  \caption{Performance of \textsc{AtomTreeSearch} for different parameter settings. The solver runs on 30 TSP instances defined on complete random Euclidean graphs for each combination of parameter values (i.e. there are 30 data points in each box of the box plot). Each row corresponds to a partitioning method. Column a) shows example partitions inducing subgraphs ${\tilde{G}}_i\left({\tilde{V}}_i,{\tilde{E}}_i\right)$ with $\max_i |\tilde{V}_i| \le M = 6$. Planar graphs are shown but the partitioning methods can be used on all types of graphs. Column b) reports the number of independent sets sampled by the QPU (\#~QPU shots) versus the TSP size when $M=\#\ qubits$ increases. Column c) shows the evolution of the optimality gap with increasing qubit count, and column d) compares performance using a quantum versus a suite of alternative classical samplers with $M=24$.}
  \label{fig:combining_algos}
\end{figure*}

\begin{figure*}[t]
  \centering
  \includegraphics[width=\textwidth]{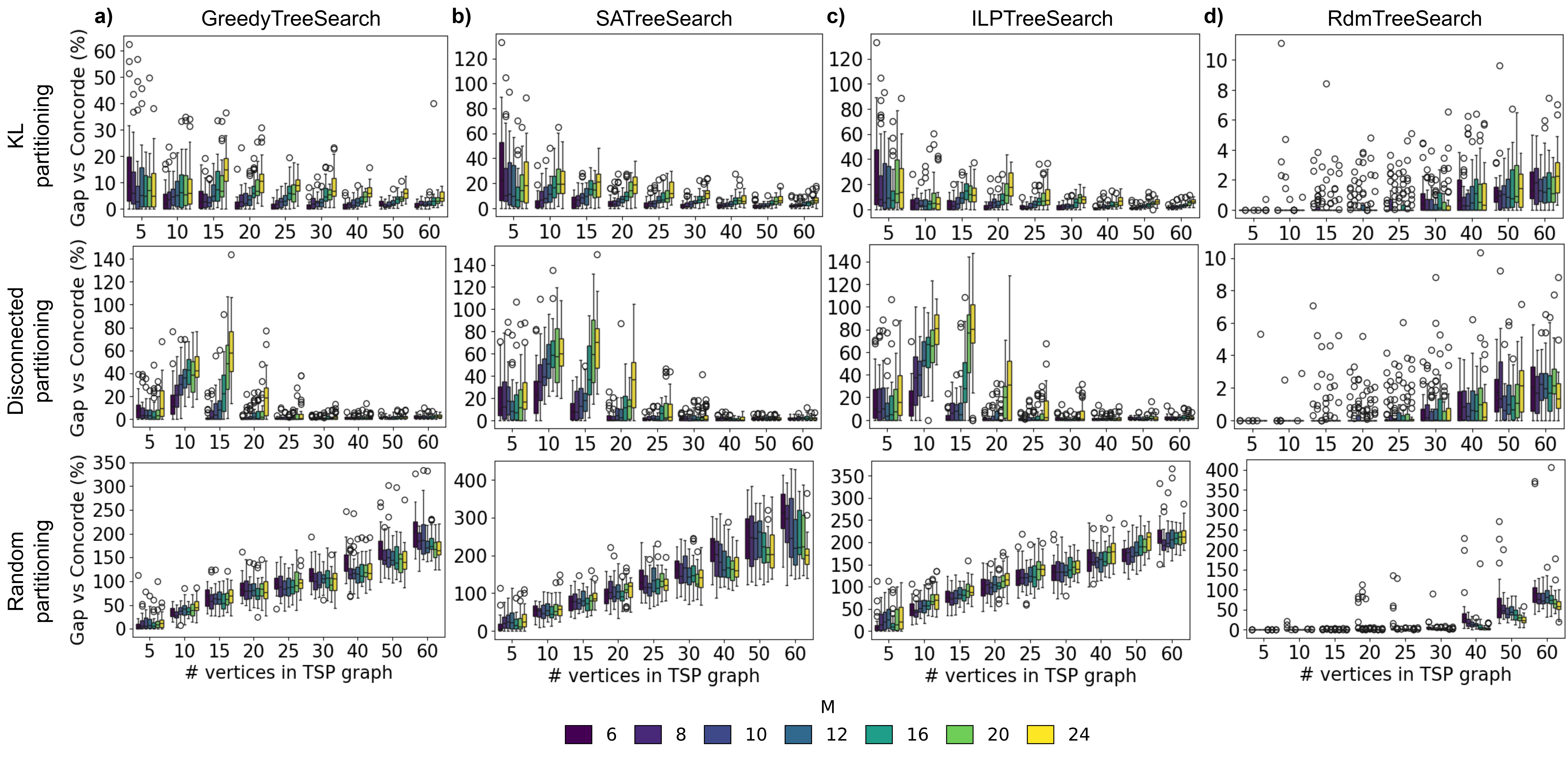}
  \caption{Performance of our tree search algorithm when using the alternative classical samplers. Each row corresponds to a partitioning method. Columns a), b), c) and d) show results for \textsc{GreedyTreeSearch}, \textsc{SATreeSearch}, \textsc{ILPTreeSearch} and \textsc{RdmTreeSearch} respectively. Results illustrate the scaling of the optimality gap obtained with these algorithms when increasing the upper bound on the size of the subgraphs $M$. The corresponding results for \textsc{AtomTreeSearch} are shown in column c) of Figure \ref{fig:combining_algos}.}
  \label{fig:class_samplers}
\end{figure*}

\begin{figure*}[t]
  \centering
  \includegraphics[width=\textwidth]{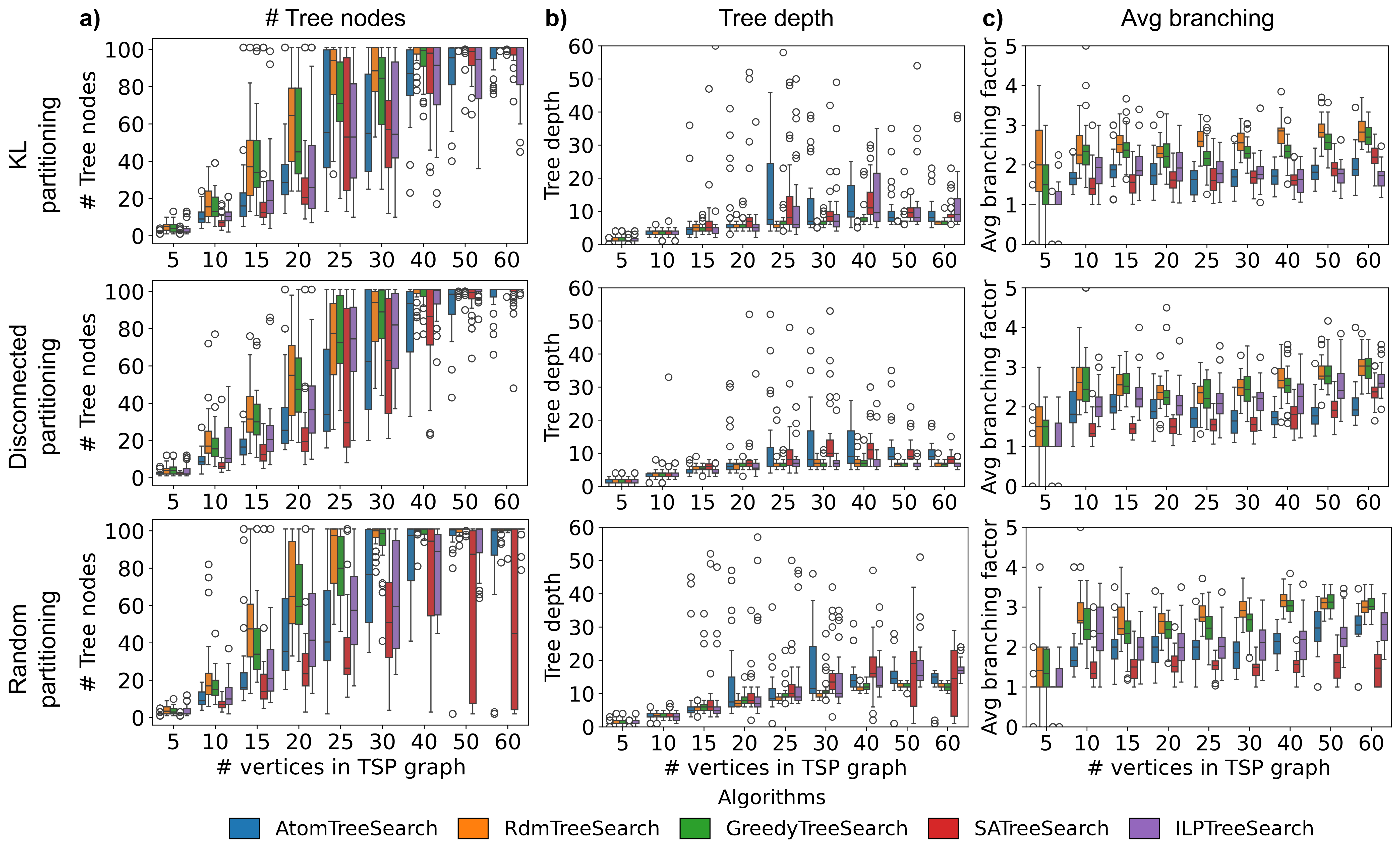}
  \caption{Comparison of the structure of trees generated when using different samplers in our tree search algorithm to solve TSP instances of varying sizes. Column a) presents the number of nodes in the trees, column b) illustrates the depth of the trees and column c) presents the average number of children a node has. Random Euclidean TSP instances with 100\% density (complete graphs) are used. The results are obtained using the KL partitioning method and $M=24$.}
  \label{fig:tree_stats}
\end{figure*}

\begin{figure*}[t]
  \centering
  \includegraphics[width=\textwidth]{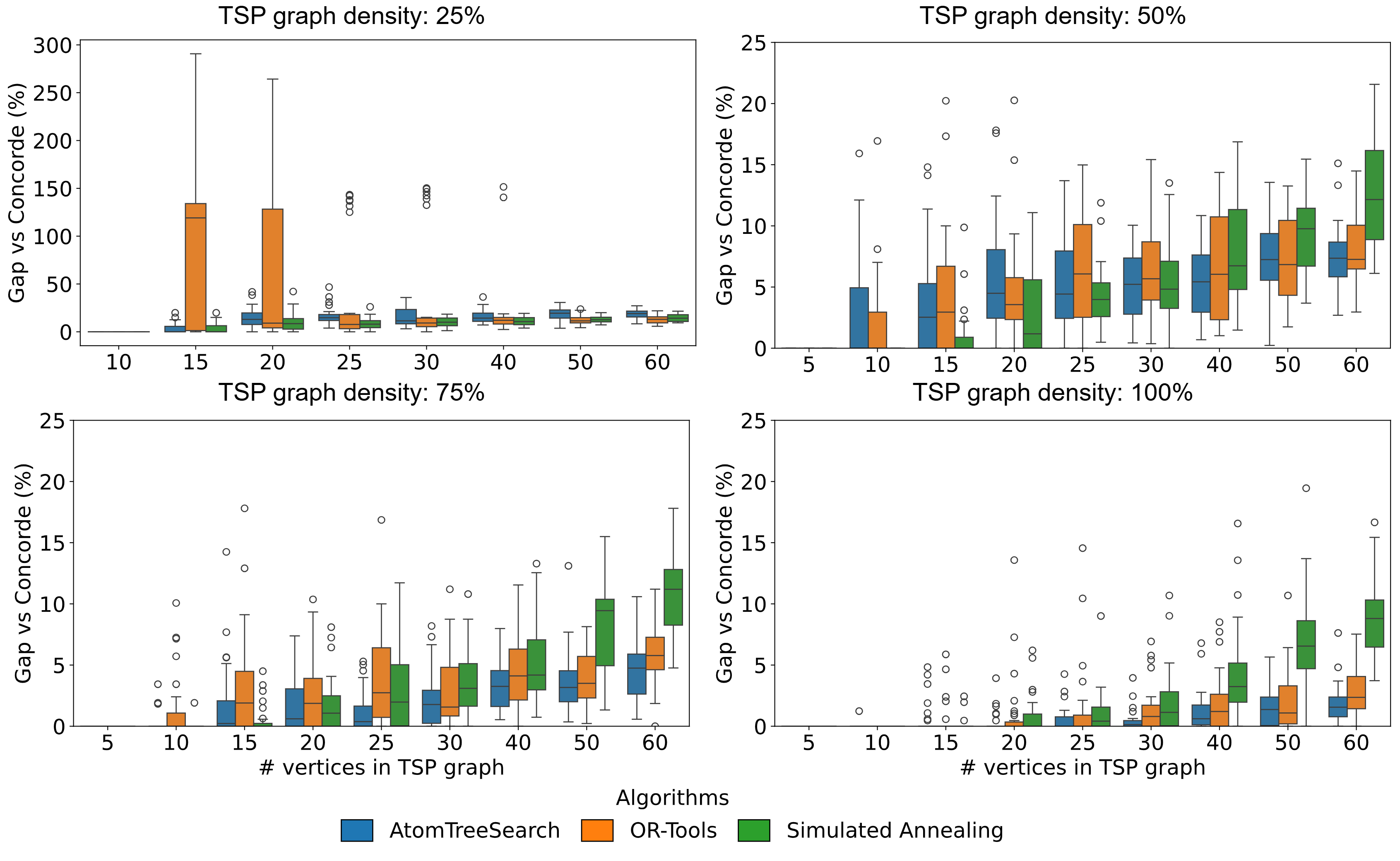}
  \caption{Performance of \textsc{AtomTreeSearch} versus \texttt{OR-Tools} and simulated annealing for different TSP sizes and different densities of the graph underlying the TSP instances. The results are obtained using the KL partitioning method and 24 qubits.}
  \label{fig:quantum_vs_class_TS}
\end{figure*}

Figures \ref{fig:combining_algos}, \ref{fig:class_samplers}, \ref{fig:tree_stats}, and \ref{fig:quantum_vs_class_TS} illustrate the performance of the previously discussed TSP solvers as various parameters are varied. Each figure is based on numerical experiments conducted on 30 TSP instances defined on random Euclidean graphs for every parameter setting, yielding 30 data points per box in the box plots. We also set the maximum number of MCTS iterations (i.e. the computational budget $b$ in Alg. \ref{alg:mcts_search}) to 100. For each MWIS solution sampler listed in section \ref{sec:solver}, the number of samples returned is set to 5 (e.g., $nb\_samples=5$ in Alg. \ref{alg:sample_mwis}). If the number of samples were increased, a corresponding increase in the maximum number of MCTS iterations would likely be required to allow the search tree to attain a comparable depth and, consequently, yield a similar level of solution quality. Furthermore, based on the analysis of Fig.~\ref{fig:tree_stats} presented in this section, we do not expect variations in the number of samples to materially affect the relative performance of the different tree-search methods. Validating this hypothesis experimentally will require substantial computational resources and will therefore be the subject of future work. Figures \ref{fig:combining_algos}, \ref{fig:class_samplers} and \ref{fig:tree_stats} present results obtained on complete graphs, whereas Figure \ref{fig:quantum_vs_class_TS} reports results for graphs with varying edge densities.

Figure \ref{fig:combining_algos} gives a first view on the performance of \textsc{AtomTreeSearch} on complete random Euclidean graphs and how this varies according to the partitioning method and the maximum allowed size of the subgraphs. The first column shows examples of subgraphs $\tilde{G_i}\left(\tilde{V_i},\tilde{E_i}\right)$ obtained with the different partitioning methods given an upper bound ($M=6$) on the size of these subgraphs. We emphasize that the partitioning methods are used before any of the quantum or classical samplers are called. This ensures a fair comparison across all samplers.

As illustrated in the first row, when using KL partitioning, we observe that the number of QPU shots to perform rapidly decreases when increasing the number of qubits as the number of subgraphs generated also decreases. Here, the number of QPU shots refers to the number of executions of lines 5-9 in Alg. \ref{alg:sample_mwis} during a single call to Alg. \ref{alg:mcts_search}. The optimality gap is generally below 2\% and decreasing with the number of qubits which is set to be equal to $M$. We note that our tree search algorithm performs significantly better with the quantum sampler than with the classical samplers except for the random sampler which provides a similar yet slightly worse performance. This behaviour is similar for different values of $M$ ranging from 6 to 24, but is slightly less pronounced for smaller values. There are evidences suggesting that \textsc{AtomTreeSearch} performs systematically better than \textsc{RdmTreeSearch} on larger quantum hardware as we shall discuss in the next paragraphs. Here we only show this comparison for $M=24$.

As shown in the third row, when using random partitioning, we observe that the number of QPU shots is significantly lower than when using the other two partitioning methods which is expected since the MWIS problem is solved for only one subgraph. For 100 iterations of the tree-search algorithm, the maximum number of QPU shots performed when using this partitioning method is 500. This upper bound is attained when each node in the search tree has at most one child. Indeed, the five QPU shots associated with the five samples returned by Alg. \ref{alg:sample_mwis} are performed only during the first execution of the \textsc{Expand} procedure on a given node. Consequently, this worst-case scenario occurs only when the 5 samples are identical for every expanded node, resulting in the creation of at most one child per node. The smaller number of QPU shots performed when using 6 or 8 qubits can be attributed to the limited number of distinct independent sets available in subgraphs containing 6 or 8 vertices. As a result, the likelihood of sampling the same independent set multiple times is relatively high, which reduces the average branching factor of the search tree and, consequently, causes the algorithm to converge faster. The optimality gap drastically reduces when increasing the number of qubits. Importantly, we observe that using the quantum sampler reduces the optimality gap signicantly compared to all other classical samplers for $M=12$ to $M=24$. Although the optimality gaps obtained with \textsc{AtomTreeSearch} and \textsc{RdmTreeSearch} are comparable for small TSP instances, a more pronounced difference emerges as the problem size increases. This observation suggests that increasing the number of qubits and the TSP instance sizes could lead to better performance from our hybrid method compared to \textsc{RdmTreeSearch}.

As shown in the second row, the use of disconnected partitioning seems to offer a compromise in balancing the number of QPU shots and the optimality gap. When this partitioning method is employed, increasing $M$ for small TSP instances reduces the number of generated subgraphs, causing the method to behave increasingly similarly to the random partitioning approach and, consequently, to achieve comparable performance. For a fixed value of $M$, this effect diminishes as the size of the TSP instances increases as expected (more subgraphs are generated). The reduced differences in tour quality achieved by the various tree-search algorithms on larger instances can be attributed to the diversity of independent sets generated for the full conflict graph through the recombination of independent sets obtained on individual subgraphs. This effect persists even when the underlying sampler yields limited solution diversity. As argued later, this increased diversity has a beneficial effect on the search process, yet purely random sampling is not enough. As shown in the third row of Fig.~\ref{fig:combining_algos}, \textsc{AtomTreeSearch}'s balance of diversity and quality leads to better results when a single subgraph is sampled.

We now turn our attention to Figure \ref{fig:class_samplers}, which illustrates the scaling of the optimality gap as a function of $M$ for the tree-search algorithms employing classical MWIS samplers. As shown in the first two rows, the optimality gap generally increases with the subgraph size $M$ when the disconnected and KL partitioning methods are used. The second row further indicates that increasing $M$ tends to degrade solution quality for small TSP instances, with the exception of \textsc{RdmTreeSearch} for reasons discussed earlier. As expected, the results presented in the third row demonstrate that the random partitioning method yields inferior performance accross all solvers, with the optimality gaps worsening as the size of the TSP instances increases.

A comparison between Figure \ref{fig:class_samplers} and column c) of Figure \ref{fig:combining_algos}, particularly with respect to the scaling of the optimality gap as a function of $M$, suggests that increasing the number of available qubits could further enhance the performance of \textsc{AtomTreeSearch} relative to its classical counterparts in terms of solution quality. Indeed, for \textsc{AtomTreeSearch}, the optimality gap either stays stable or decreases with the number of qubits (size of subgraphs) as illustrated by column c) of Figure \ref{fig:combining_algos}. Furthermore, the number of QPU shots decreases with qubit count when using KL partitioning, and in a less pronounced way when using disconnected partitioning, as fewer subgraphs are generated. On the other hand, the sampling task gets computationally harder for classical methods as these subgraphs get larger. We also note that partitioning methods tend to become equivalent as $M$ gets larger. As $M$ tends to infinity, all three methods generate only one subgraph equal to the original graph. To reiterate, given all these observations, we expect the performance gap between the hybrid method and the classical ones to widen with qubit count regardless of the partitioning method used.

We now turn to a study of the structure of the trees generated when using the different samplers. Figure \ref{fig:tree_stats} illustrates the number of tree nodes, the tree depth and the average number of children a node has when using the different samplers. The variations in tree structure observed in Fig.~\ref{fig:tree_stats} do not provide sufficient evidence to explain alone the differences in optimality gap achieved by the various tree-search algorithms presented in Figure \ref{fig:combining_algos} d). Therefore, we conclude that the observed performance differences are related to the behavior of the samplers in a more nuanced manner, which is not fully captured by the tree-structure metrics considered here. In what follows, we nevertheless provide our interpretation of the mechanisms that may underlie the differences in tree structure and how these mechanisms may affect performance.

We observe that the random and greedy samplers tend to generate search trees containing a larger number of nodes and characterized by greater width and lower depth than those produced by the other three samplers. This suggests that these samplers generally return a diverse set of independent sets, thereby encouraging the tree-search algorithm to explore a broader portion of the solution space. The performance difference between \textsc{RdmTreeSearch} and \textsc{GreedyTreeSearch} can, in turn, be attributed to differences in the distribution of solution quality, with the greedy sampler producing independent sets whose qualities are more homogeneous.

In contrast, the simulated annealing sampler produces trees with fewer nodes, greater depth, and a smaller branching factor. This behavior suggests that the sampler frequently returns the same independent sets for a given node, resulting in fewer child nodes and consequently reducing the extent of exploration performed by the tree-search algorithm.

The integer linear programming sampler generally yields trees whose average branching factor lies between those obtained with the greedy and simulated annealing samplers. By construction, the ILP sampler returns the best independent sets, implying a very low diversity in solution quality among the sampled solutions, even if the solutions themselves are diverse (e.g., a large Hamming distance between bitstrings). This limited variability may reduce differences in the exploitation term of Eq. \ref{UCT}, thereby causing the tree-search algorithm to place greater emphasis on exploration.

Finally, the quantum MWIS sampler returns independent sets with probabilities that are approximately proportional to their quality. As a result, it appears to reach a favorable balance between solution quality and solution diversity. This balance is beneficial to the classical tree-search procedure, leading to deeper search trees than those produced by \textsc{RdmTreeSearch} while maintaining a substantial degree of exploration of the solution space.

Regarding the relative performance of \textsc{AtomTreeSearch} compared to classical TSP solvers, Figure \ref{fig:quantum_vs_class_TS} shows that, for dense random Euclidean graphs, our hybrid algorithm is competitive with both \texttt{OR-Tools} and simulated annealing alone even when given limited resources (24 qubits). On the other hand, as graphs become sparser when we remove available edges to traverse, all three solvers perform similarly. We expect that this is due to the limited number of feasible Hamiltonian cycles in sparse graphs. The lack of local structure connecting two Hamiltonian cycles is likely keeping all methods stuck in the same region of the solution space.

\begin{figure*}[t]
  \centering
  \includegraphics[width=\textwidth]{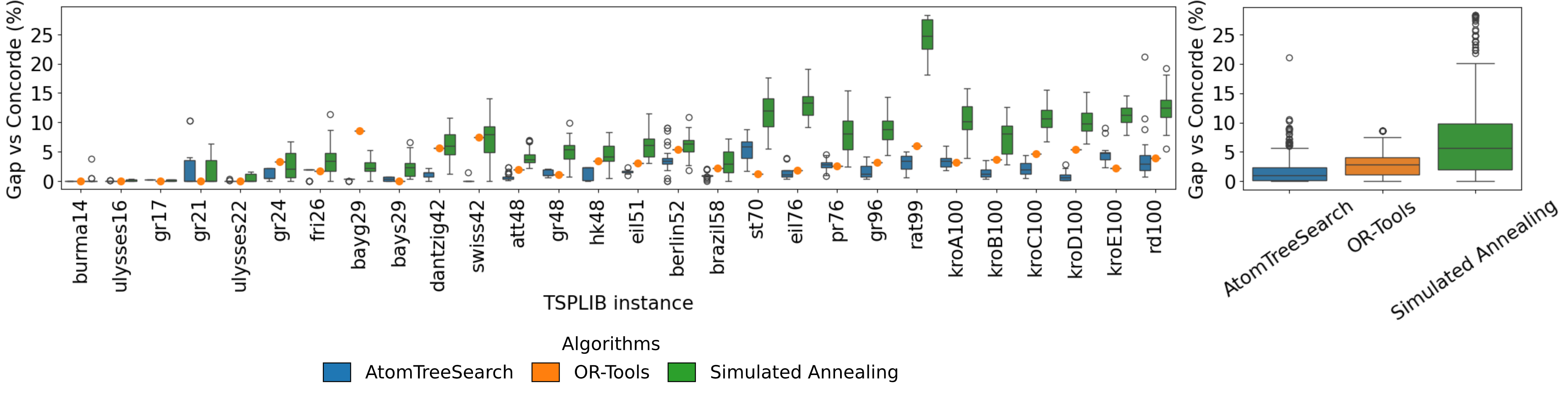}
  \caption{Performance of \textsc{AtomTreeSearch} compared with \texttt{OR-Tools} and simulated annealing on TSPLIB instances with up to 100 cities. Because \textsc{AtomTreeSearch} and simulated annealing are stochastic algorithms, each solver is run 30 times per instance, resulting in a distribution of optimality gaps, which is summarized using box plots. The left panel shows the performance on individual instances, while the right panel presents the aggregated performance across all instances. The results are obtained using the KL partitioning method and 24 qubits. \textsc{AtomTreeSearch} is given a budget of 100 MCTS iterations.}
  \label{fig:TSP_solvers}
\end{figure*}

When benchmarked on TSPLIB instances with up to 100 cities, \textsc{AtomTreeSearch} with a budget of 100 MCTS iterations generally matches or outperforms \texttt{OR-Tools} and simulated annealing, as illustrated in Figure \ref{fig:TSP_solvers}. The performance of \textsc{AtomTreeSearch} on this practical benchmark set is consistent with the trends observed for complete graphs in Fig.~\ref{fig:quantum_vs_class_TS}. In particular, simulated annealing appears to be largely unable to effectively explore the solution space of larger instances when relying exclusively on successive 2-Opt moves. By contrast, the collective updates performed by \textsc{AtomTreeSearch} enable radical tour reconfigurations, allowing the algorithm to find better overall solutions. Moreover, \textsc{AtomTreeSearch} exhibits lower variability in solution quality than simulated annealing, despite both methods being fundamentally stochastic. \texttt{OR-Tools}, while deterministic, produces inferior tours compared to those obtained with \textsc{AtomTreeSearch} in the majority of the tested instances. Finally, we note that \textsc{AtomTreeSearch} can temporarily leave the feasible solution space when collective actions break the Hamiltonian cycle. This capability may be advantageous, as it allows the search process to traverse infeasible intermediate states and potentially converge to higher-quality solutions compared to simulated annealing and \texttt{OR-Tools}. Together, these results demonstrate the effectiveness of \textsc{AtomTreeSearch} as a quantum-guided heuristic for combinatorial optimization problems.

\section{Conclusion}\label{sec:conclusion}
In this paper, we presented a new quantum-enhanced MCTS framework for combinatorial optimization problems called \textsc{AtomTreeSearch}. We explained how a MWIS problem can be introduced in the usual UCT algorithm and how a quantum sampler running on a neutral-atom quantum computer can provide solutions to this problem. Indeed, at each expansion step, a maximal weighted independent set of candidate actions provided by the quantum processor is selected, and these collective actions are performed to obtain a child node. When adapting \textsc{AtomTreeSearch} to specific combinatorial optimization problems, one need only specify a set of individual actions that one can take to go from one solution to another and the independence relation between these actions. We considered the case of the TSP, a paradigmatic optimization problem with well-honed solvers available for benchmarking. In this setting, individual actions can be taken as 2-Opt operations. Our analysis shows that the integration of a simple quantum subroutine based on a quantum annealing protocol significantly helps the tree search. We also found that on random Euclidean instances with up to 60 cities and TSPLIB instances with up to 100 cities, our hybrid algorithm generally matches or outperforms both \texttt{OR-Tools} and a simulated annealing approach.

Since current neutral-atom quantum processors provide far more qubits than those used in the limited simulations presented in this work \citep{Pasqal_Docs_2025}, the proposed algorithm is executable today on existing hardware, where we expect to be able to tackle much larger subgraphs. As neutral-atom quantum computers counting thousands of atoms and shot frequency in the thousands of Hz are not so far away in the future \citep{Pichard_2024}, we expect that it will be possible to use our algorithm within a few years on large TSP instances that are hard to approximate for competing classical heuristics. The results we show in Fig. \ref{fig:TSP_solvers} comparing the performance of \textsc{AtomTreeSearch} on TSPLIB instances with that of \texttt{OR-Tools} and a simulated annealing approach point in that direction already, even for smaller problems and under limited computational resources due to our inability to emulate our quantum protocol at a larger scale.

Our algorithm can be used to tackle other combinatorial optimization problems in the routing and logistics area. In particular, it is applicable to various forms of the VRP, including the asymmetric TSP, capacitated VRP, VRP with time windows and other constrained VRP variants for which the 2-Opt operator remains applicable. Other straightforward adaptations of our algorithm are possible to solve other combinatorial optimization problems by defining different actions and the concept of independence between these actions. Inspiration can often be taken from improvement heuristics applied to these problems, as they define improvement operations that can be taken as actions in the context of our algorithm.

Both our framework and our implementation for the TSP can be improved in multiple ways, encompassing both its classical and quantum components. First, as proposed in SP-MCTS by \cite{Schadd2008}, the UCB1 tree policy could be augmented with a deviation term, and a meta-search strategy periodically restarting the search with a different root could be applied. Also, as pointed out by \cite{Bjornsson2009}, since averaging simulation results can hide small regions of good solutions if the surrounding solutions are weak, we could keep track of maximum simulation results at each node to guide the search. Moreover, although we employed MCTS in an offline fashion, it is generally used as an online planning algorithm. While this choice may be adequate for small TSP instances, larger instances might benefit from an online planning approach. Alternative partitioning and kernelization strategies could replace the KL, disconnected, and random methods used here. The simulation step itself might also be improved through the use of quantum-enhanced Markov Chain Monte Carlo methods \citep{Layden2023}, for example.

On the quantum side, we employed both a relatively simple quantum annealing protocol and a straightforward embedding strategy, but several alternative approaches could be considered. With respect to pulse shaping, as discussed by \cite{Leclerc_2024} and \cite{dalyac_2023}, the time-dependent profiles of the laser Rabi frequency and detuning could be optimized using techniques such as Bayesian optimization \citep{Frazier_2018}, stochastic gradient descent \citep{Mason_1999} and methods from quantum optimal control such as the Krotov method \citep{Krotov1993}, gradient ascent pulse engineering (GRAPE) \citep{Khaneja2005} and chopped random basis (CRAB) optimization \citep{Caneva2011}. Machine learning methods also present opportunities as discussed by \cite{coelho2022} and \cite{Vercellino_2022} for example. \cite{coelho2022} also propose several force-directed embedding methods that could be used in \textsc{AtomTreeSearch}. In addition, \cite{perron2026leveraging} introduce an algorithm based on simulated annealing to address the embedding problem, which represents another promising alternative. Also, \cite{Zhang2026} demonstrated that an analog counterdiabatic quantum computing protocol can be used to solve MWIS problems on neutral-atom quantum computers and obtain encouraging results. Such a protocol could be employed in our framework to obtain higher quality samples.

Regarding our implementation for the TSP, tighter bounds such as the Held–Karp lower bound and improved heuristics for upper bounds could enhance performance. Although there are $O(N^2)$ and $O(N^3)$ 2-Opt and 3-Opt operations respectively for a given tour, the 2-Opt and 3-Opt algorithms can run in subquadratic time in practice by relying on techniques such as neighbor lists and ``don't look bits,'' as discussed by \cite{johnson2002}. Incorporating these ideas could therefore accelerate the process of defining a new conflict graph for each node. Additionally, 3-Opt operations could be considered when the conflict graph becomes sparse at deeper tree levels and other k-Opt operations could be adaptively added based on the sparsity of the currently obtained conflict graph. Finally, a different heuristic could be used during the simulation phase. For example, simulated annealing could be replaced with a tabu-search approach, to approach a Lin--Kernighan style search. Collectively, these improvements may position \textsc{AtomTreeSearch} as a hybrid quantum-classical heuristic with performances comparable to state-of-the-art methods for a range of industrially relevant combinatorial optimization problems.

\section*{Acknowledgment}
This research was financially supported by the Natural Sciences and Engineering Research Council of Canada (NSERC) and Pasqal Canada through an Alliance grant (Alliance 59081-2023). This work made use of computer resources by Calcul Québec and the Digital Research Alliance of Canada.

\appendix
\renewcommand{\thesection}{Appendix \Alph{section}}
\section{Software versions}

\setlength{\tabcolsep}{23pt}
\begin{table}[H]
    \centering
    \begin{tabular}{ll}
         \hline
         Software&Version \\
         \hline
         Python&3.11.4\\
         QuTiP&5.1.1 \\
         EMU-SV&2.3.0 \\
         EMU-MPS&2.3.0 \\
         Pulser&1.6.1 \\
         OR-Tools&9.6.2534 \\
         Concorde&03.12.19 \\
         \hline
    \end{tabular}
    \caption{Sofware versions used in this work.}
    \label{tab:placeholder}
\end{table}

\bibliography{bib}

\end{document}